\documentclass[%
 aip,
 amsmath,amssymb,
 reprint,%
]{revtex4-1}

\usepackage{graphicx}% Include figure files
\usepackage{dcolumn}% Align table columns on decimal point
\usepackage{bm}% bold math
\usepackage[utf8]{inputenc}
\usepackage[T1]{fontenc}
\usepackage{mathptmx}
\usepackage{etoolbox}

\makeatletter
\def\@email#1#2{%
 \endgroup
 \patchcmd{\titleblock@produce}
  {\frontmatter@RRAPformat}
  {\frontmatter@RRAPformat{\produce@RRAP{*#1\href{mailto:#2}{#2}}}\frontmatter@RRAPformat}
  {}{}
}%
\makeatother
\begin{document}

\preprint{AIP/123-QED}

%\title[Damped Oscillations of a Biomimetic Scale-covered Viscoelastic Substrate]{Damped Oscillations of a Biomimetic Scale-covered Viscoelastic Substrate}

\title{Material-Geometry Interplay in Damping of Biomimetic Scale Beams}%{Damped Oscillations of a Biomimetic Scale-covered Viscoelastic Substrate}
% Force line breaks with \\

\author{H. Ebrahimi}
%\email{ebrahimi@knights.ucf.edu}
\affiliation{Department of Mechanical and Aerospace Engineering, University of Central Florida, 4000 Central Florida Blvd, Orlando, FL 32816}%Lines break automatically or can be forced with \\
\author{M. Krsmanovic}%
%\email{milos@knights.ucf.edu}
\affiliation{Department of Mechanical and Aerospace Engineering, University of Central Florida, 4000 Central Florida Blvd, Orlando, FL 32816}%Lines break automatically or can be forced with \\
\author{H. Ali}
%\email{alih@union.edu}
\affiliation{Stress Engineering Services, Inc., 7030 Stress Engineering Way, Mason, OH 45040}%Lines break automatically or can be forced with \\
%\author{P. Warren}
%\email{peterwarren@knights.ucf.edu}
%\affiliation{Department of Mechanical and Aerospace Engineering, University of Central Florida, Orlando, FL}%Lines break automatically or can be forced with \\
\author{R. Ghosh}%
\email{ranajay.ghosh@ucf.edu}
\affiliation{Department of Mechanical and Aerospace Engineering, University of Central Florida, 4000 Central Florida Blvd, Orlando, FL 32816}%Lines break automatically or can be forced with \\

\date{\today}% It is always \today, today,

\begin{abstract}
Biomimetic scale-covered substrates are architected meta-structures exhibiting fascinating emergent nonlinearities via the geometry of collective scales contacts. In spite of much progress in understanding their elastic nonlinearity, their dissipative behavior arising from scales sliding is relatively uninvestigated in the dynamic regime. Recently discovered is the phenomena of viscous emergence, where dry Coulomb friction between scales can lead to apparent viscous damping behavior of the overall multi-material substrate. In contrast to this structural dissipation, material dissipation common in many polymers has never been considered, especially synergestically with geometrical factors. This is addressed here for the first time, where material visco-elasticity is introduced via a simple Kelvin-Voigt model for brevity and clarity. The results contrast the two damping sources in these architectured systems: material viscoelasticity, and geometrical frictional scales contact. It is discovered that although topically similar in effective damping, viscoelsatic damping follows a different damping envelope than dry friction, including starkly different effects on damping symmetry and specific damping capacity.
\end{abstract}

\maketitle

\begin{figure*}
 \centering
 \begin{tabular}{cc}
 \includegraphics[width=3.1in]{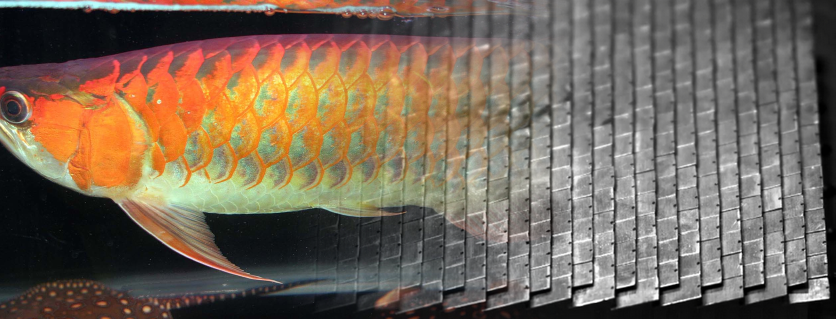}
 & \includegraphics[width=3.1in]{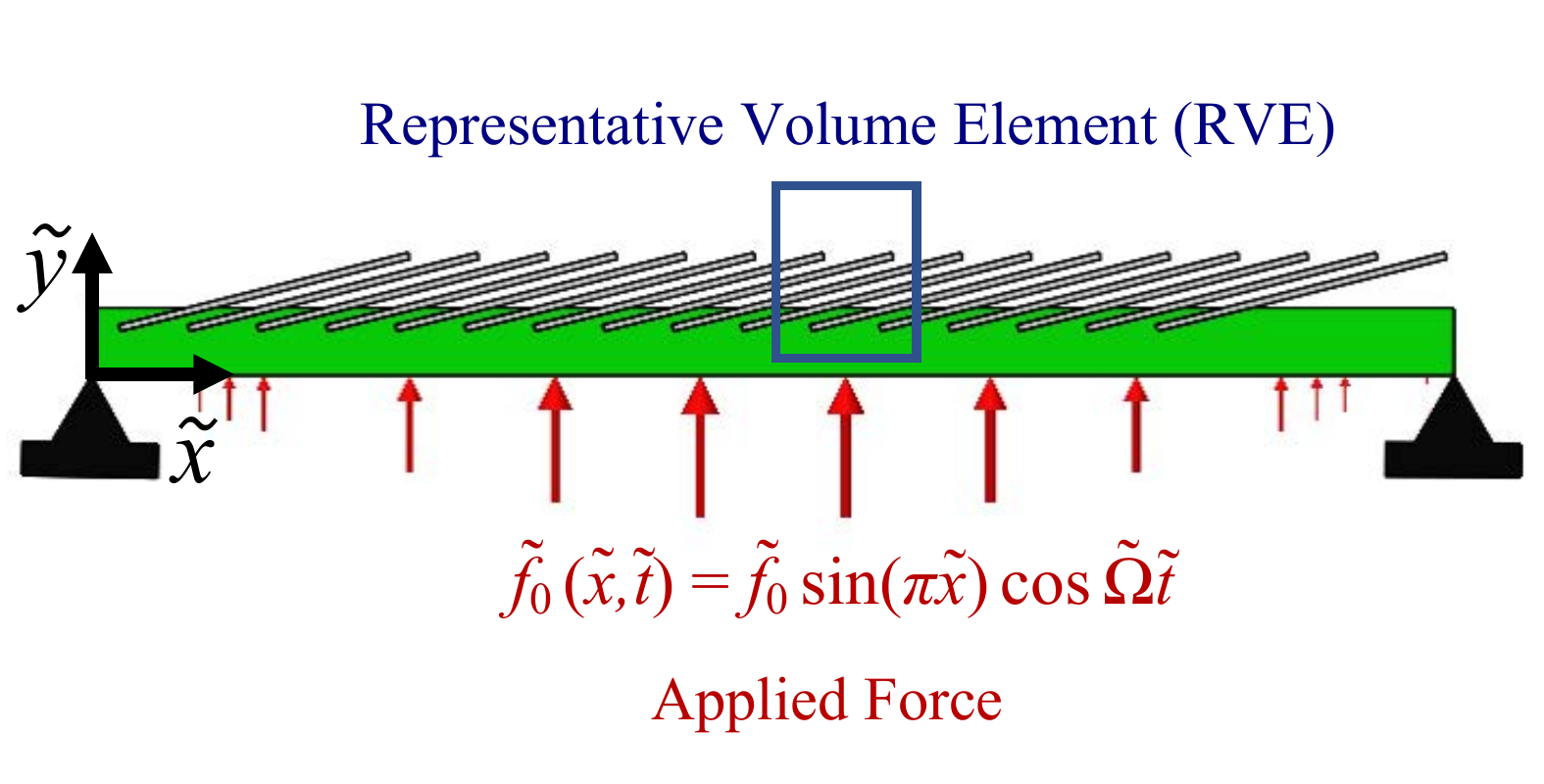} \\
 (a) & (b) \\ 
 \includegraphics[width=3.1in]{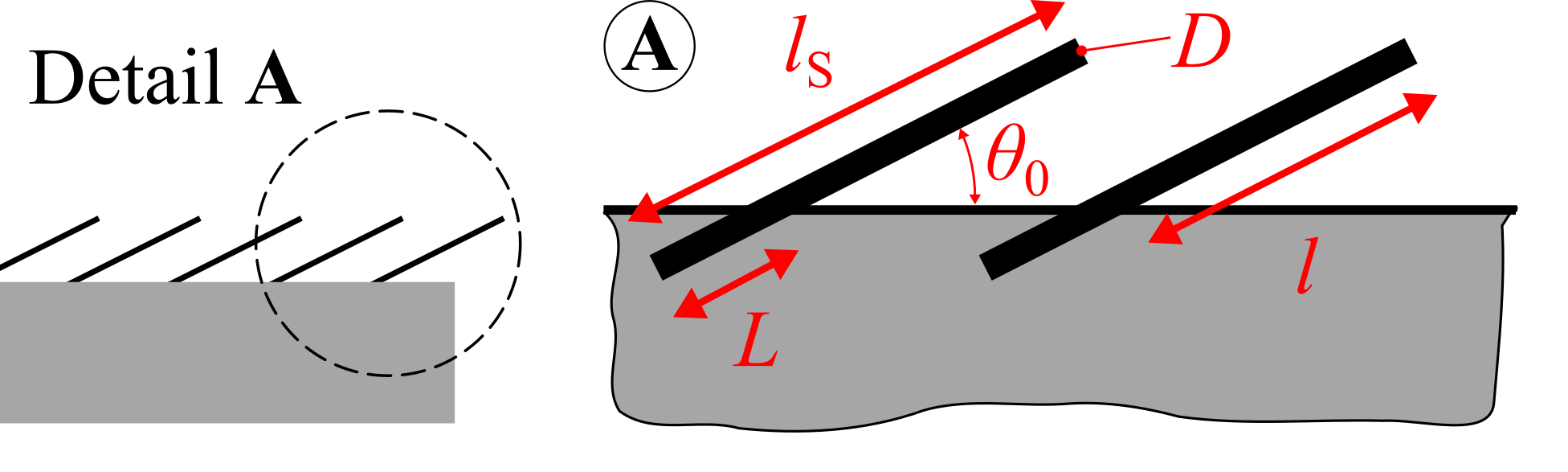}
 & \includegraphics[width=3.1in]{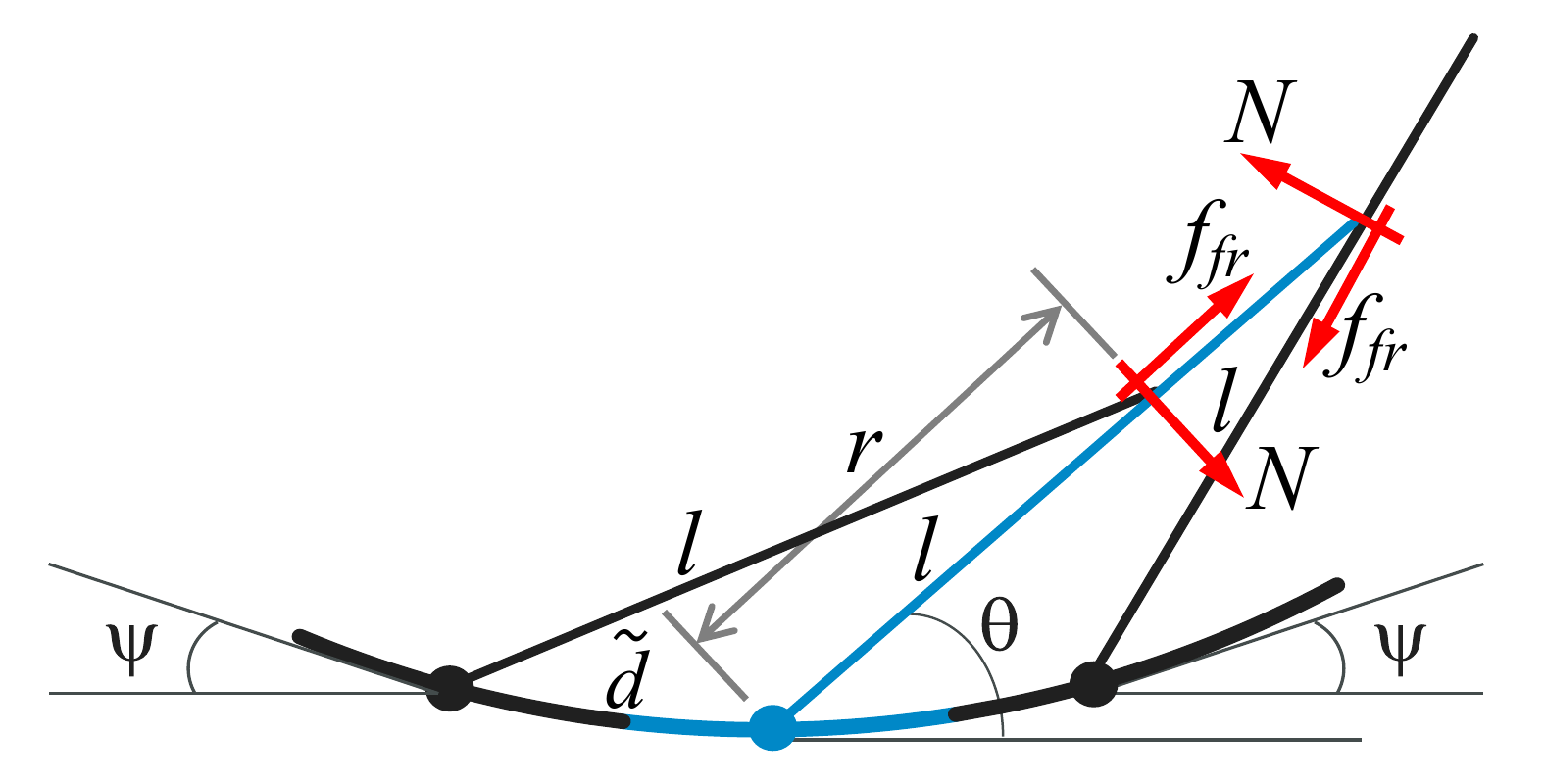} \\
 (c) & (d) \\
\end{tabular}
 \caption{(a) The overlapped arrangement of natural fish scales, and their artificial biomimetic reproduction. The picture is adapted under CC BY 2.0 \cite{aini2008sting}. (b) Schematic diagram of a simply supported biomimetic scaled beam and the representative volume element (RVE) selected from the middle of the beam marked with the blue rectangle. (c) Schematic diagram of scales with characteristic dimensions and angles. (d) Schematic geometry of the representative volume element (RVE) with detailed geometric parameters.}
 \label{Fig1}
 \vspace{10pt}
\end{figure*}

Biologically inspired scale-covered substrates are under sustained scrutiny as a structural platform with unique property combinations akin to metamaterials using the geometry and kinematics of scales sliding Fig. \ref{Fig1} \cite{natarajan2022performance, muthuramalingam2020transition, jiakun2021review, rose2021biomimetic, jenett2017digital}. Physically, such multi-material systems comprise of a soft deformable substrate with protruding stiff plates acting as scales,  
%When the substrate deforms, the scales eventually contact in a collective giving rise to a fascinating spectrum of elastic properties[cite]. In quasi-static loading, these include for instance a nonlinear strain stiffening even in small strains, instantaneous interlocking state, emergent anisotropy in bending, twisting and indentation, and regime-differentiated elasticity. Scaly substrates also display unique fracture and failure behavior.  Many of these behaviors emerge from the collective sliding motion of the scales on the substrate. Such kinematic origins of nonlinear behavior mean that their fundamental origins lie in the distribution of the scales. This results in a fascinating geometry-dictated landscape of nonlinear elasticity and fracture. [Need citations]
Fig. \ref{Fig1} (a) \cite{long1996functions,aini2008sting}. %Physically, such multi-material systems comprise a soft deformable substrate with protruding plate-like stiff parts that act as scales \cite{browning2013mechanics,ghosh2014contact,ebrahimi2019tailorable,ali2019bending,ali2019frictional}. 
When the substrate deforms, the stiff scales eventually contact as a collective giving rise to a fascinating spectrum of mechanical and optical properties \cite{ghosh2014contact,ebrahimi2019tailorable,ali2019bending}. %In quasi-static loading, these include nonlinear strain stiffening even in small strains, instantaneous interlocking states, emergent anisotropy in bending, twisting, and indentation, and a regime-differentiated elasticity \cite{ghosh2014contact,ghosh2016frictional,ali2019bending,ali2019frictional,ali2020tailorable,ebrahimi2019tailorable,ebrahimi2020coulomb,ebrahimi2021emergent,ebrahimi2021fish,dharmavaram2022coupled,ali2019tailorable}. Scale-covered substrates also display unique fracture and failure behavior \cite{lin2011mechanical, liu2016numerical}. 
Many of these behaviors emerge from the collective sliding motion of the scales on the substrate. Such kinematic origins of nonlinear behavior mean that their fundamental source of nonlinearity lie in the distribution and orientation of the stiff scales \cite{wang2016pangolin,yang2013natural, ehrlich2015materials}. This results in a geometry-dictated landscape of nonlinear elasticity and fracture. 

In spite of deep scrutiny of the elastic and fracture characteristics, interest in the dissipative behavior of these substrates has been more recent \cite{ghosh2016frictional,ali2019frictional,ebrahimi2020coulomb,dharmavaram2022coupled}. In the static regime where dry friction was postulated between scales, friction was found to play a dual role in adding stiffness to the substrate as well as limiting the range of motion by introducing an additional locking phase \cite{ghosh2016frictional,ebrahimi2020coulomb,ali2020tailorable}. Such conflicting roles prompted further extension of friction onto the dynamic regime. Here, a new type of dissipative behavior was discovered - emergent viscosity. In other words, even when dry friction was assumed between the scales, the overall damped oscillation of the substrates indicated viscous-like exponential damping \cite{ali2019frictional}. 

In these works, the role of substrate polymer viscoelasticity was not investigated \cite{ali2019frictional}. Thus, the only source of damping was from the scales sliding. However, many polymers exhibit viscoelastic behavior \cite{ferry1980viscoelastic,polacco2006relation}, and its interplay with scales sliding dissipation remains unknown. Specifically, the effect of damping that emerges from the synergistic combination of the viscoelasticity of the substrate and the dry interfacial scale friction has not been revealed. In this letter, we include the viscoelasticity of the substrate for the first time to understand its role of damping during the oscillation of an Euler-Bernoulli substrate. For this study, we chose a simple Kelvin-Voigt model to represent the viscoelasticity \cite{craifaleanu2015bending,freundlich2019transient}, and assume Coulomb friction between the scales. We investigate both free vibration and forced vibration.

% For the structure of a biomimetic scale beam, we consider a viscoelastic substrate with rigid rectangular partially embedded scales. To describe the viscoelastic behavior of the substrate, the Kelvin–Voigt model has been considered for the viscoelastic material properties. 
Viscoelasticity of real polymers is a highly complex phenomenon encompassing both linear and nonlinear deformations and multiple intrinsic time scales. In this letter, a simple Kelvin-Voigt model is chosen for this study for brevity and fundamental understanding of the complex geometry-material interplay with the aim for further detailed numerical studies in later publications. We also note that the major source of nonlinearity in these slender substrates come from geometrical and contact sources and not the material due to the geometrical locking phenomena \cite{ghosh2014contact}. The Kelvin-Voigt model for viscoelastic behavior can be represented as a purely viscous damping element (damper) and a purely elastic element (spring) connected parallel together. The relationship between stress $\sigma$, strain $\varepsilon$, and strain rate $\mathrm{d}\varepsilon/\mathrm{d}t$ is governed by \cite{craifaleanu2015bending,freundlich2019transient}:
\begin{equation}\label{Eq1}
    \sigma = E_B \varepsilon + \tilde{\xi}\frac{\mathrm{d}\varepsilon}{\mathrm{d}t},
\end{equation}
where $E_B$ and $\tilde{\xi}$ are the Young's modulus and viscosity of the substrate, respectively. %In this 1D biomimetic scale-covered beam, the scales are inclined with $\theta_0$ as the initial scales' inclination angle, and partially embedded on the substrate's top surface, with uniform spacing along the length of the beam as shown in Fig. \ref{Fig1} (c). As mentioned earlier,% 
The scales are considered rigid to isolate the purely geometric effect of scales.
%, and this rigidity assumption is valid because the scales in the real fish skin and also in the fabricated samples of biomimetic scales metamaterials are several orders of magnitude stiffer in comparison with the substrate material \cite{ghosh2014contact,vernerey2010mechanics}.

The length of substrate is considered as $L_B$, and the height $h_B$. We match the material properties of a typical silicone rubber, which can be used to fabricating the soft substrate. These material properties are as follows: the Young's modulus $E_B=1.5$ MPa, Poisson's ratio $\nu=0.42$ \cite{ali2019bending,ali2019tailorable}, and density $\rho_B=854$ kg/m$^3$ \cite{ali2019frictional}. The viscosity of silicone rubbers is in the range of 1 to $10^8$ mPa.s \cite{MatWeb2022,suriano2020viscoelastic}, where the lower range is related to the liquid form of silicone polymer and the higher range is related to the solid form of silicone polymer. %Then, we averagely consider the viscosity of substrate as $\tilde{\xi}=0.015$ MPa.s in this paper. 
It should be noted that these material properties are just an example of a soft viscoelastic silicone polymer, and their exact values are not critical to the central discoveries of this research work.

%In this problem, in addition to the material nonlinearity due to the viscoelasticity, the nonlinearity is also stemming from scales interaction. 
The rectangular scales with thickness $D$ are partially embedded into the top surface of the substrate with initial angle $\theta_0$, Fig. \ref{Fig1} (c). The total length of scales $l_s$ is including the exposed length $l$, and embedded length $L$, ($l_s=l+L$). The exposed length of scales $l$ is non-dimensionalized by the spacing between the neighboring scales $\tilde{d}$, , Fig. \ref{Fig1} (c), as $\eta=l/\tilde{d}$ called overlap ratio \cite{ghosh2014contact,ali2019frictional}. We assume that the scale’s thickness $D$ is negligible in comparison with the length of the scales, $l_s$ ($D \ll l_s$), and the scale’s embedded length is negligible in comparison with the substrate’s height ($L \ll h_B$) \cite{ghosh2014contact,ghosh2016frictional,ebrahimi2019tailorable,ebrahimi2020coulomb,ali2019frictional,ali2019bending}. These assumptions allow us to consider each scales as a linear torsional spring with constant $\tilde{K}_s$, due to resistance of substrate against rotation of embedded part of scales   \cite{ghosh2014contact,vernerey2010mechanics,ali2019bending}. For such a system, the constant was obtained as scaling expression $\tilde{K}_s=E_B D^2 C_B (L/D)^n$, where $C_B$ and $n$ are constants with corresponding values $0.66$ and $1.75$, respectively \cite{ghosh2014contact, ali2019frictional}.

The dynamic equation of motion of a viscoelastic plain beam (without any scale) is derived using Hamilton’s principle, $\delta \int_{\tilde{t}_1}^{\tilde{t}_2}{(\hat{T}-\hat{V}+W})d\tilde{t} = 0$ for a viscoelastic beam. In this relationship, $\hat{T}$, $\hat{V}$, and $W$ are the kinetic energy per unit length, the strain energy per unit length, and the work done by the applied traction, respectively, \cite{javadi2021nonlinear,ali2019frictional}--which leads to the following differential equation for a plain viscoelastic beam \cite{craifaleanu2015bending,diani2020free}:

\begin{equation}\label{Eq2}
    {\rho _B}A_B{\frac{{\partial ^2}\tilde y} {\partial {{\tilde t}^2}}} + {E_B}{I_B}{\frac{{\partial ^4}\tilde y}{\partial {{\tilde x}^4}}} + \tilde{\xi} I_B \frac{\partial}{\partial \tilde{t}} \Big(\frac{\partial^4 \tilde{y}}{\partial \tilde{x}^4}\Big) = \tilde{f}( {\tilde x,\tilde t} ),
    \vspace{6pt}
\end{equation}
Here, $A_B$ and $I_B$ are the area of the substrate's cross-section and the second moment of area, respectively. The quantities: $\tilde{t}$, and $\tilde{x}$ and $\tilde{y}$ are time, and the two spatial coordinates shown on Fig. \ref{Fig1} (b), respectively. The $\tilde{f}( {\tilde x,\tilde t})$ is the applied force function, which is shown in Fig. \ref{Fig1} (b) schematically, for a scale-covered beam. For the free vibration case, the function $\tilde{f}( {\tilde x,\tilde t})$ is equal to zero and the equation of motion is a homogeneous equation, whereas for the forced vibration case, applied force can be considered as the first mode $\tilde{f}(\tilde{x},\tilde{t}) = \tilde{f}_0 \phi(\tilde{x})\cos{\tilde{\Omega}\tilde{t}}$ \cite{abhyankar1993chaotic,ali2019frictional}, where $\tilde{f}_0$, $\phi(\tilde{x})$, and $\tilde{\Omega}$ are the load amplitude, the first mode shape function for the simply supported beam, and the load frequency \cite{rao2019vibration,ali2019frictional}. First mode shape function for a simply supported beam with length $L_B$ is known as $\phi(\tilde{x})=\sin \big(\frac{\pi \tilde{x}}{L_B}\big)$. % In forced vibration, the equation of motion is a non-homogeneous equation (because $\tilde{f}(\tilde{x},\tilde{t}) \neq 0$ in Eq. (\ref{Eq2})), and the general solution $\tilde{y}(t)$ can be given by the sum of homogeneous solution $\tilde{y}_h(t)$, and the particular solution $\tilde{y}_p(t)$ as follows $\tilde{y}(t)=\tilde{y}_h(t)+\tilde{y}_p(t)$. The homogeneous solution $\tilde{y}_h(t)$ (the free vibration part) dies out with time if there is a damping source in the system, and the general solution eventually reduces to the particular solution $\tilde{y}_p(t)$. This means the forced vibration system after a transient motion reaches a steady-state vibration \cite{rao2019vibration}. 
%To this end, the damping source in the system can be provided by either the artificial viscous damping $\tilde{C}$ (aka air damping) or the viscoelasticity of the beam $\tilde{\xi}$ to stabilize the numerical solution for the steady state forced oscillation cases. In this paper, we consider the substrate as a viscoelastic material ($\tilde{\xi}>0$), then we can neglect the air damping ($\tilde{C}=0$).

% Now, we consider a uniform distribution of partially embedded rigid rectangular plates as biomimetic scale on top of the substrate along its length, as shown in Fig. \ref{Fig1} (c). 1D biomimetic scaled beam with uniform spacing can preserve periodicity under pure bending through scales engagement, which means all scales start to engage at the same time with similar configuration. 
Under pure bending, the relationship between scales inclination angle $\theta$, and the substrate bending angle $\psi$, is given as $\theta = \sin^{-1}(\eta\psi \cos\psi/2)-\psi/2$ \cite{ghosh2014contact,ghosh2016frictional,vernerey2010mechanics}. Note that although this relationship is not satisfied globally except pure bending, local periodicity could be assumed (dense scales assumption) \cite{ali2019frictional, ali2019bending}.

With these considerations, the global deformation of the scaly beam can be envisioned as a combination of the two deformation modes comprising of the substrate bending and the local scales rotating in all RVEs \cite{ghosh2014contact,ali2019bending}. This kinematics allows the inclusion of the work of the friction between scales as they slide. The frictional work can be included in the Hamilton's principle now. 
%By deriving the variational form of the energy equation using Hamilton’s principle, the dynamic equation of motion of the scale-covered viscoelastic beam can be obtained, based on the kinetic and strain energy of the substrate per unit length, dissipation energy due to the substrate's viscoelasticity and air damping per unit length, the strain energy per unit length due to the rotation of partially embedded rigid scales, the dissipation energy due to the friction between the scales, and the work done by the applied traction, \cite{javadi2021nonlinear,ali2019frictional}. 
The friction is modeled based on the Coulomb dry friction by considering different coefficients of friction $\mu$, and the effect of scales' mass on the kinetic energy of the system is neglected \cite{ali2019frictional}. These considerations lead to the following partial differential equation for a scale-covered viscoelastic beam \cite{ali2019frictional} (see Supplementary Material): 

\begin{eqnarray}
\label{Eq3}
  {\rho _B}A_B{\frac{{\partial ^2}\tilde y}{\partial {{\tilde t}^2}}} + {E_B}{I_B}{\frac{{\partial ^4}\tilde y}{\partial {{\tilde x}^4}}} + \tilde{\xi} I_B \frac{\partial}{\partial \tilde{t}} \Big(\frac{\partial^4 \tilde{y}}{\partial \tilde{x}^4}\Big)
 + {\frac{{\partial ^2}}{\partial {{\tilde x}^2}}}\hspace{-1pt}{1 \over {{N}}}\hspace{-1pt} \sum\hspace{-1pt} \bigg{[} \hspace{-1pt} {{\tilde K}_s} ( {{\theta} - {\theta _0}} )  \nonumber\\
 {{\partial {\theta}} \over {\partial {\psi}}} + {{\sin (\beta){{\tilde K}_s}( {{\theta} - {\theta _0}} ){\rm{sgn}}({\mathop {\dot{\tilde y}}\limits^{} })} \over {\cos ( {{\psi } + \beta } ) - {\bar r}\cos ( \beta  )}} {{\partial {\bar r}} \over {\partial {\psi }}}\bigg{]} H( {{{\tilde \kappa }} - {{\tilde \kappa }_e}} ) = \tilde{f}( {\tilde x,\tilde t} ). \hspace{15pt}
\end{eqnarray}
\vspace{5pt}

Here, $\beta=\tan^{-1}{\mu}$, and $\bar{r}$ is the non-dimensionalized form of $r$, which is the distance between the scale's base to the interaction point with the left neighboring scale, with respect to the exposed length of scale $l$, as $\bar{r}=r/l$, Fig. \ref{Fig1} (d) for a particular RVE. By considering the geometrical arrangements in each RVE shown in Fig. \ref{Fig1} (d) $\bar{r}$ is derived as $\bar{r}=\frac{\sin \big(\theta-\psi/2\big)}{\sin \big(\theta+\psi/2\big)}$. In Eq. (\ref{Eq3}), the Heaviside step function ensures that the terms regarding the strain energy due to the scales rotation, and the dissipation energy due to the friction between the scales, are only contributed after scales engagement at each RVE level. That is, only in the case of downward deflection of the beam, and when $\theta > \theta_0$ or, in another word, when $\tilde{\kappa}>\tilde{\kappa}_e$. The number of RVEs utilized in the solution of the system has been shown as $N$. 

% The non-dimensionalization of Eq. (\ref{Eq3}) has been described in Appendix \ref{sec:appendixA}. After non-dimensionalization, this equation is solved through a computational analysis using the Newmark beta scheme as a direct numerical integrator, which uses a Newton-Raphson nonlinear equation solver \cite{ahmadian2012non}.

\begin{figure}
 \centering
 \includegraphics[width=3.4in]{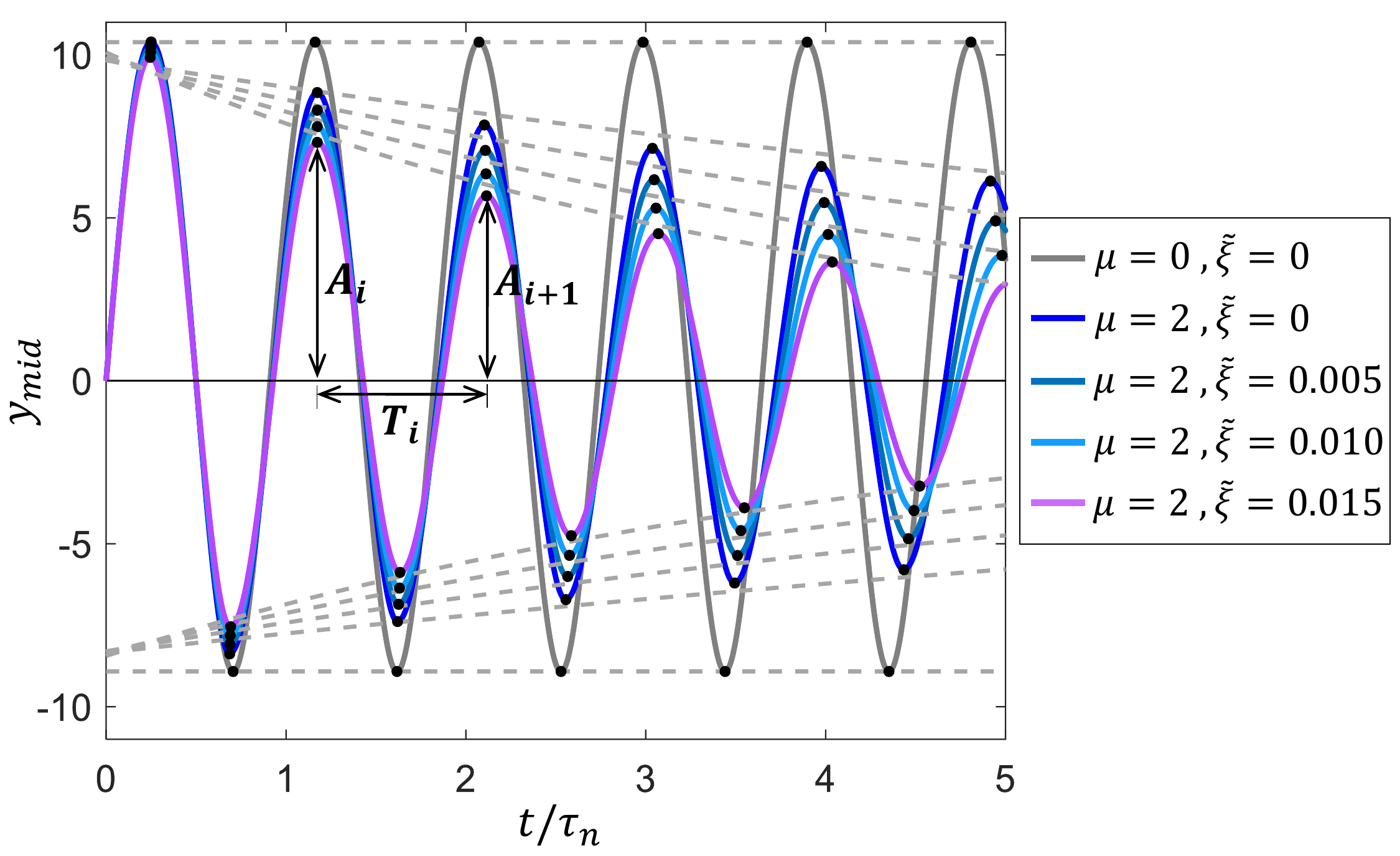}
\caption{The response of middle point of the scaly viscoelastic beam with $\eta=5$ under velocity initial condition for various viscosity coefficients $\tilde{\xi}$ (Different $\tilde{\xi}$ are described with the unit MPa.s). For the gray plot $\mu=0$, and for other plots $\mu=2$.}
\label{Fig2}
\end{figure}

We first verify our model by comparing the midpoint deflection during free vibration of the beam with finite element simulation of an equivalent system, Fig.~\ref{Fig2}. In this figure, which is a displacement-time plot of the midpoint of the beam, we fix $\eta=5$ , $\theta_0=5^\circ$ and vary the coefficient of friction. The black dots indicate FE simulations, which are in excellent agreement with our model results. %These plots highlight the effect of increasing the material viscosity $\bar\xi$ of the underlying substrate. 
Overall, comparing this plot with purely Coulombic friction case~\cite{ali2019frictional}, it looks as if the material viscosity effects are very similar to dry friction effect. They both lead to viscous damping, and increase with time. %and also an asymmetry of vibration amplitude on the two sides of the beam.
Thus, it would seem that dry friction and material viscosity effects reinforce in tandem, the viscous damping of the beam. 

However, this is where their similarities end. The material viscosity is essentially an symmetric source of dissipation - acting on both sides of the bending whereas the friction is asymmetric, acting only on scales side. In addition, the scales themselves add asymmetry to the overall vibration by dent of being on only one side.  %However, the anisotropy in vibration brought about by scales themselves are generally low until much higher density of overlap~\cite{ali2019frictional}. Thus, it can be expected that higher initial scales angles that lead to later engagement would have a symmetric influence on the overall behavior due to reduced region of engagement. However, this is not entirely true as frictional forces are known to increase exponentially with curvature~\cite{ghosh2016frictional} and thus much more effective in the later stages of deformation. Thus, the effect of both material parameters and geometry have dual roles to play in vibration - together by reinforcing the damping characteristic and against each other in breaking symmetry of vibration.  

The asymmetry brought about by scales have a pronounced geometrical component. On the one hand, if the scales are dense, the additional stiffness would be higher on the scales side. Similarly, frictional effects would also be higher. Thus it would seem like denser scales add to asymmetry of the medium. On the other hand, if the scales initial inclination is higher, then they engage at a greater curvature and hence their impact on symmetry would be lesser. The effect of material viscoelasticity seems to be symmetric in nature because it acts on both sides. However, the scales on one side also inhibit displacement on the other side, and hence the symmetric effect of material viscosity can also be broken. A suitable measure of asymmetry would be the logarithmic decrement factor ($\delta=\frac{1}{\Delta}\log\frac{AA_{n+1}}{AA_n}$) that measures the relative decline of amplitudes in successive cycles. This parameter is related to the overall damping coefficient and the Q-factor of the vibration. We could expect that these vibration asymmetries would lead to a split in the $\delta$ values between the concave and convex side. We define the ratio of the two $\delta$s by an asymmetry ratio $\alpha=\delta_{convex}/\delta_{concave}$, and take them as a measure of bi-directional asymmetry. 

\begin{figure*}
 \centering
\begin{tabular}{cc}
 \includegraphics[width=2.7in]{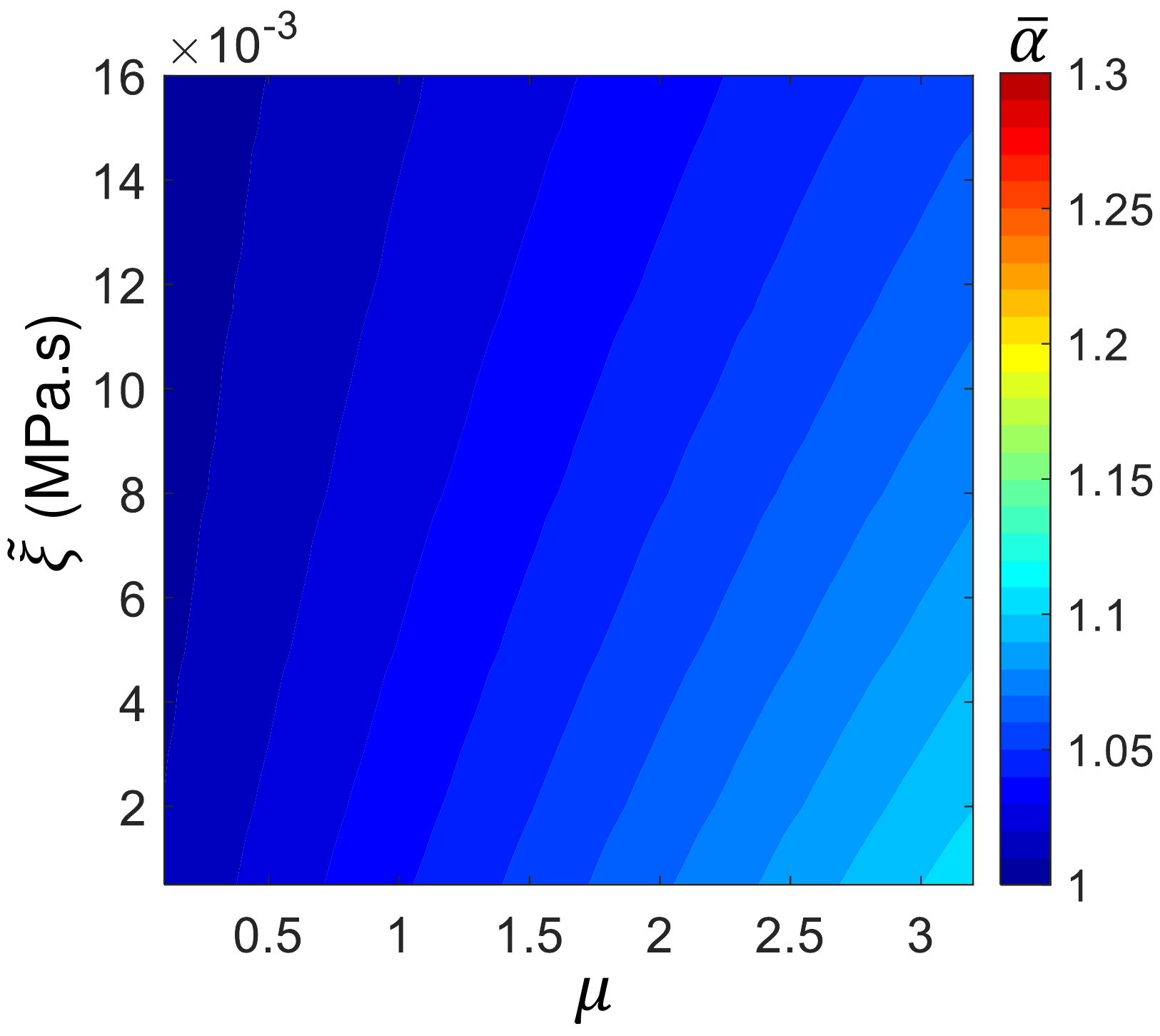} 
 % & \includegraphics[width=2.3in]{Fig3(b).pdf}
 & \includegraphics[width=2.7in]{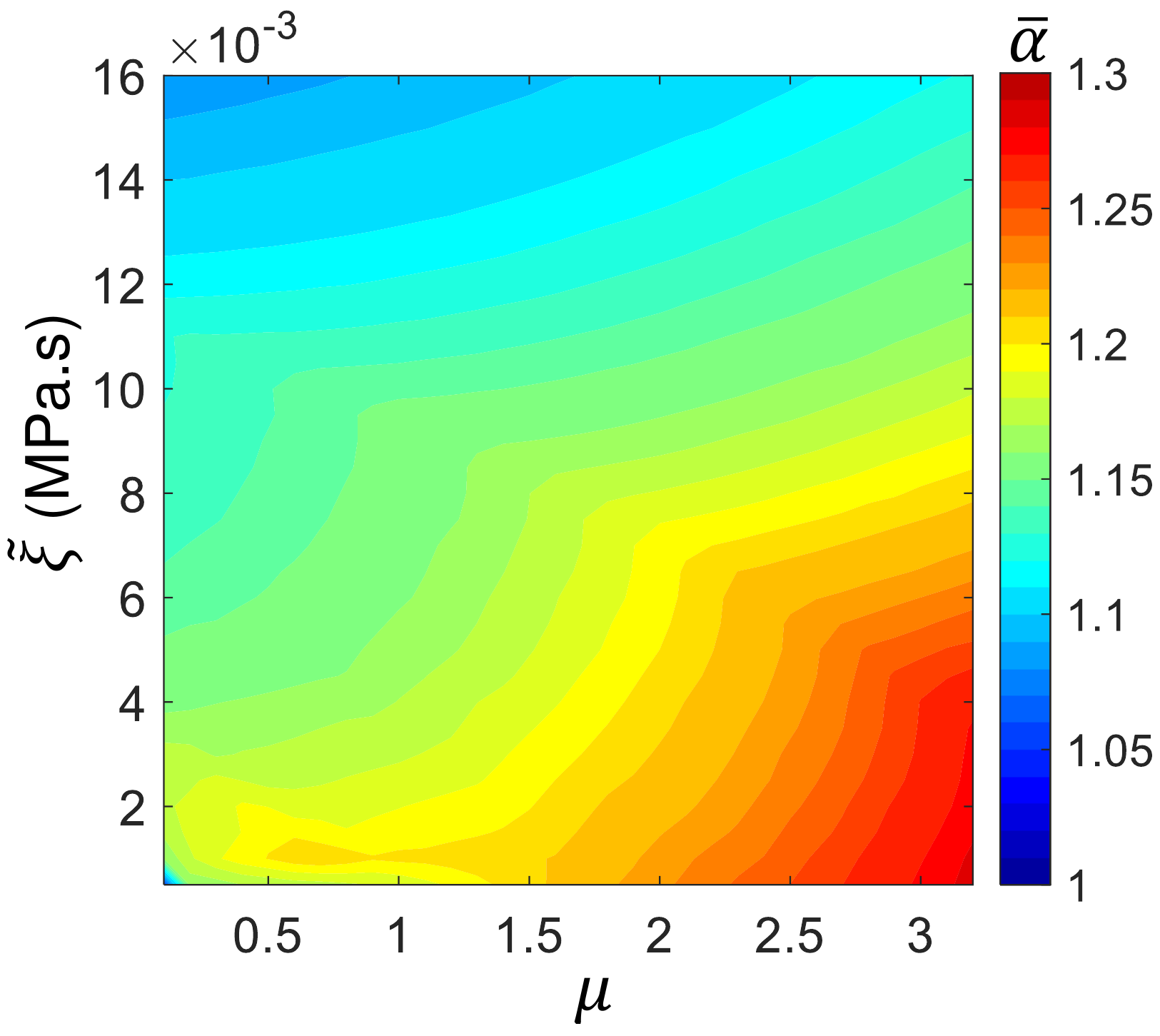}\\
 (a) & (b) \\% & (c)\\
  \includegraphics[width=2.7in]{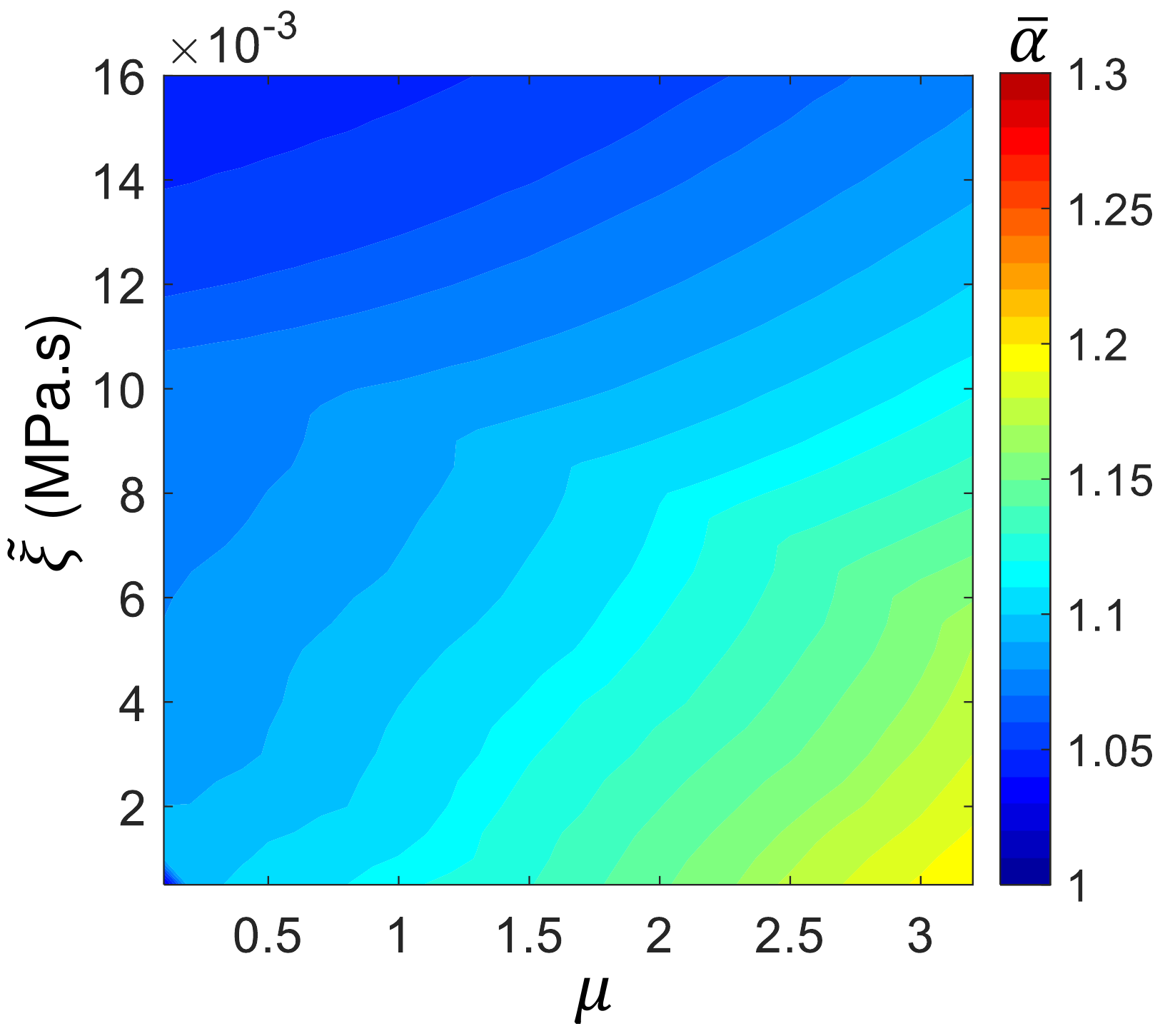} 
 & \includegraphics[width=2.7in]{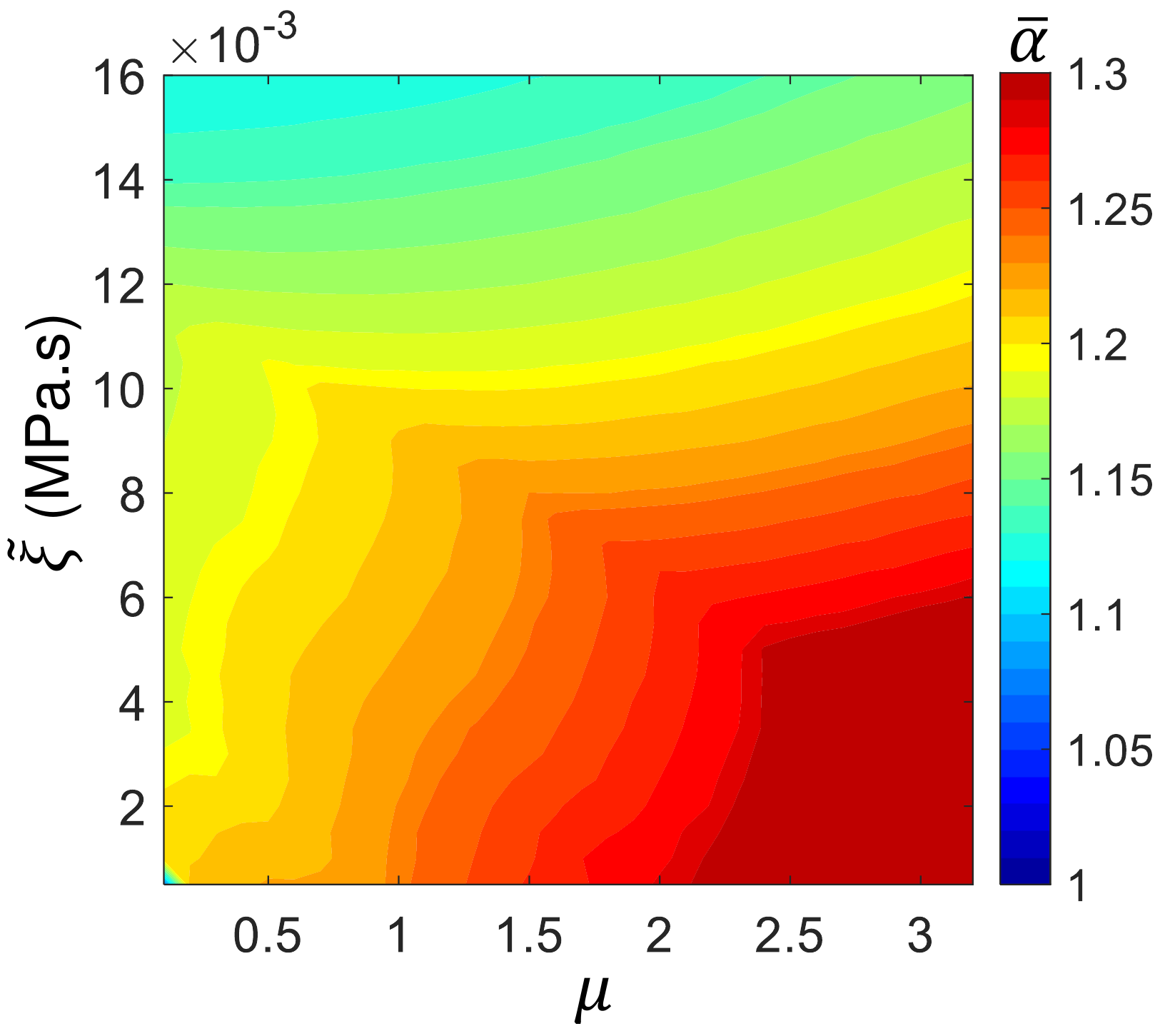}\\
 (c) & (d) \\ % & (f) \\
\end{tabular}
\caption{Phase map of the ratio between the convex and concave damping coefficient
($\bar{\alpha}$), known as asymmetry coefficient, spanned by $\tilde{\xi}$ and $\mu$, for different cases: (a) $\eta=5$, and $\theta_0=0^\circ$. (b) $\eta=5$, and $\theta_0=5^\circ$. (c) $\theta_0=5^\circ$, and $\eta=3$. (d) $\theta_0=5^\circ$, and $\eta=7$.}
\label{Fig3}
\end{figure*}

%To investigate this anisotropy, we use the logarithmic decrement defined as $\delta={\frac{1}{\Delta}}\ln(AA_{n+1}/AA_n )$ where $\Delta$ is the time between $n$ and $n+1$ peaks for both sides of oscillations. We then take a ratio of the two to quantify the amount of bi-directional asymmetry, $\alpha =\delta_{convex}/\delta_{concave}$. 
Topically, it would seem that increasing friction of scales would cause greater asymmetry as it acts selectively on only one direction whereas material dissipation would cause dissipation symmetrically in both direction. Hence, increasing Coulomb friction should accentuate asymmetry, whereas viscoelasticity should leave it unaffected. In order to investigate these effects, we develop phase maps of $\alpha$ mapped by $\mu$ and $\bar\xi$ for various values of $\eta$ and $\theta_0$, Fig.~\ref{Fig3}. The first row of this asymmetry map Fig.~\ref{Fig3} (a-b) shows the effect of increasing $\theta_0$ while $\eta$ is kept constant. Fig.~\ref{Fig3} (a) shows that increasing Coulomb friction does not lead to an increase in anisotropy of logarithmic damping, even though its effect on displacement asymmetry is pronounced. A rather surprising and counter-intuitive result. It seems like symmetry is broken only when the initial inclination angle changes. Once that occurs, increasing inclination angle causes the asymmetric region to flatten and spread to lower values of Coulomb friction. We also investigate the effect of scale density $\eta$ towards the asymmetry, Fig.~\ref{Fig3} (c-d). As expected the effect of higher density is also to further anisotropy. For the same combination of $\mu$ and $\bar \xi$, the anisotropy is greatly pronounced with higher $\eta$, Fig.~\ref{Fig3} (c-d).

\begin{figure*}
 \centering
\begin{tabular}{cc}
 \includegraphics[width=2.7in]{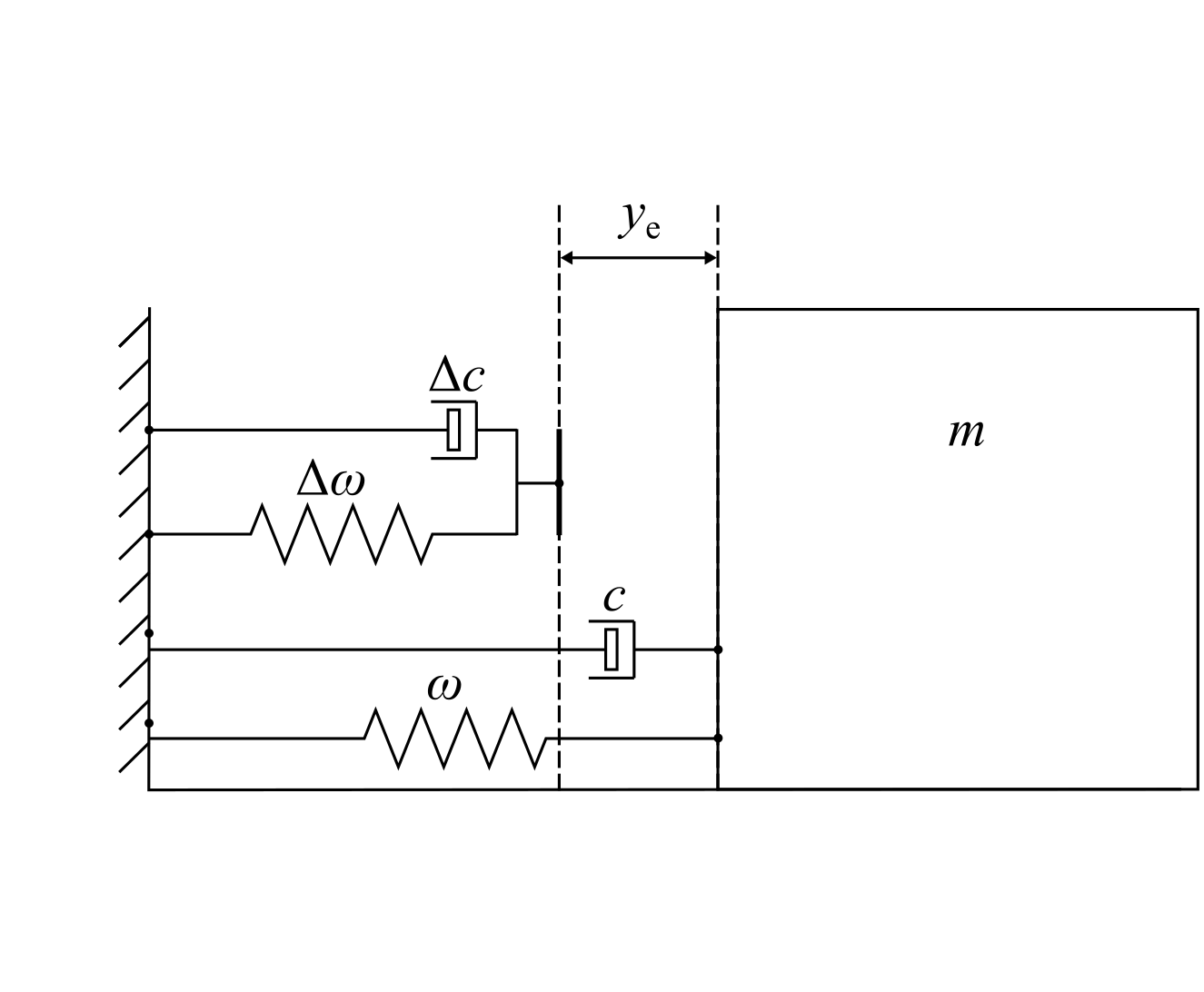} 
 & \includegraphics[width=2.7in]{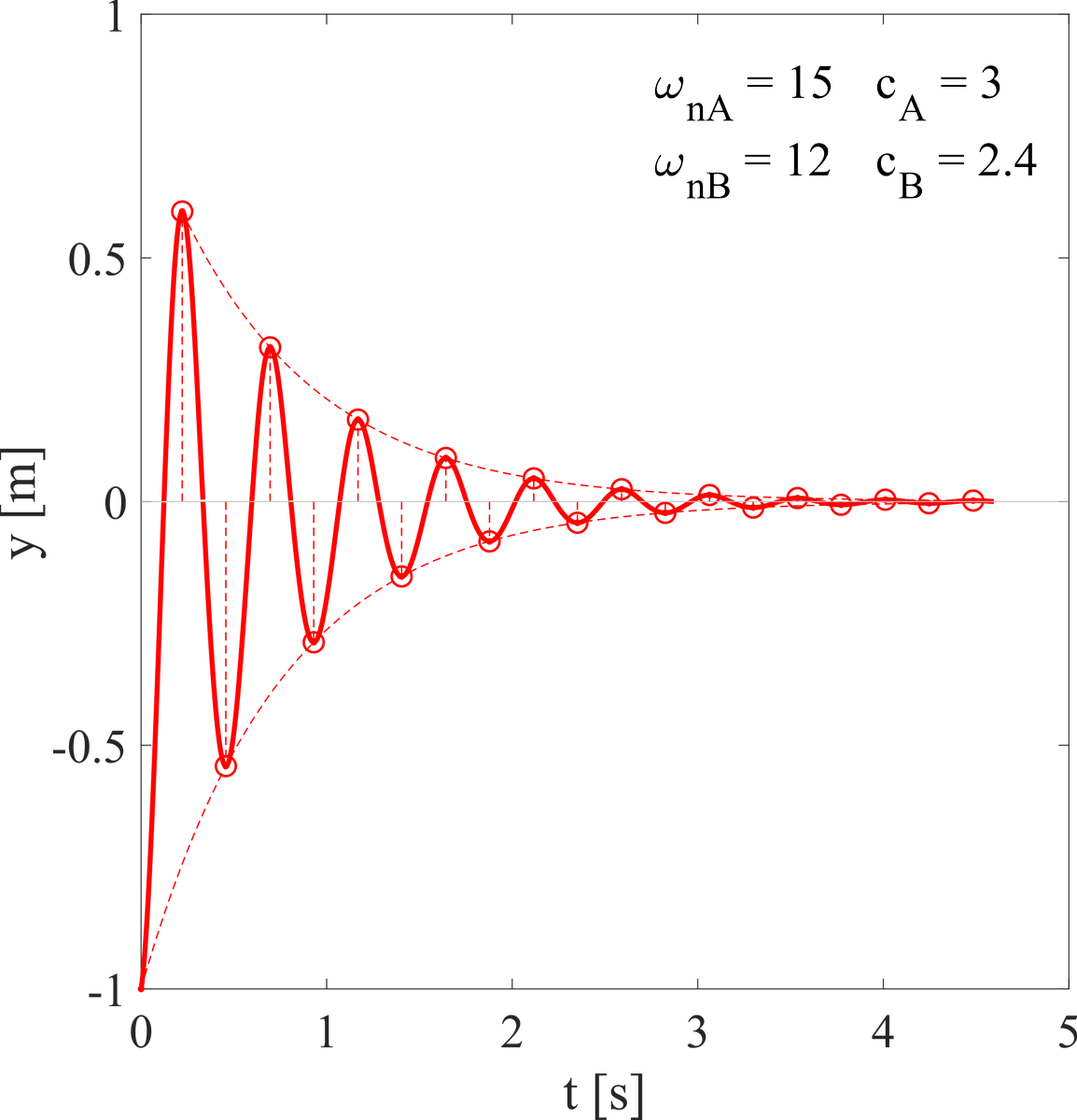}\\
 (a) & (b) \\
 \includegraphics[width=2.7in]{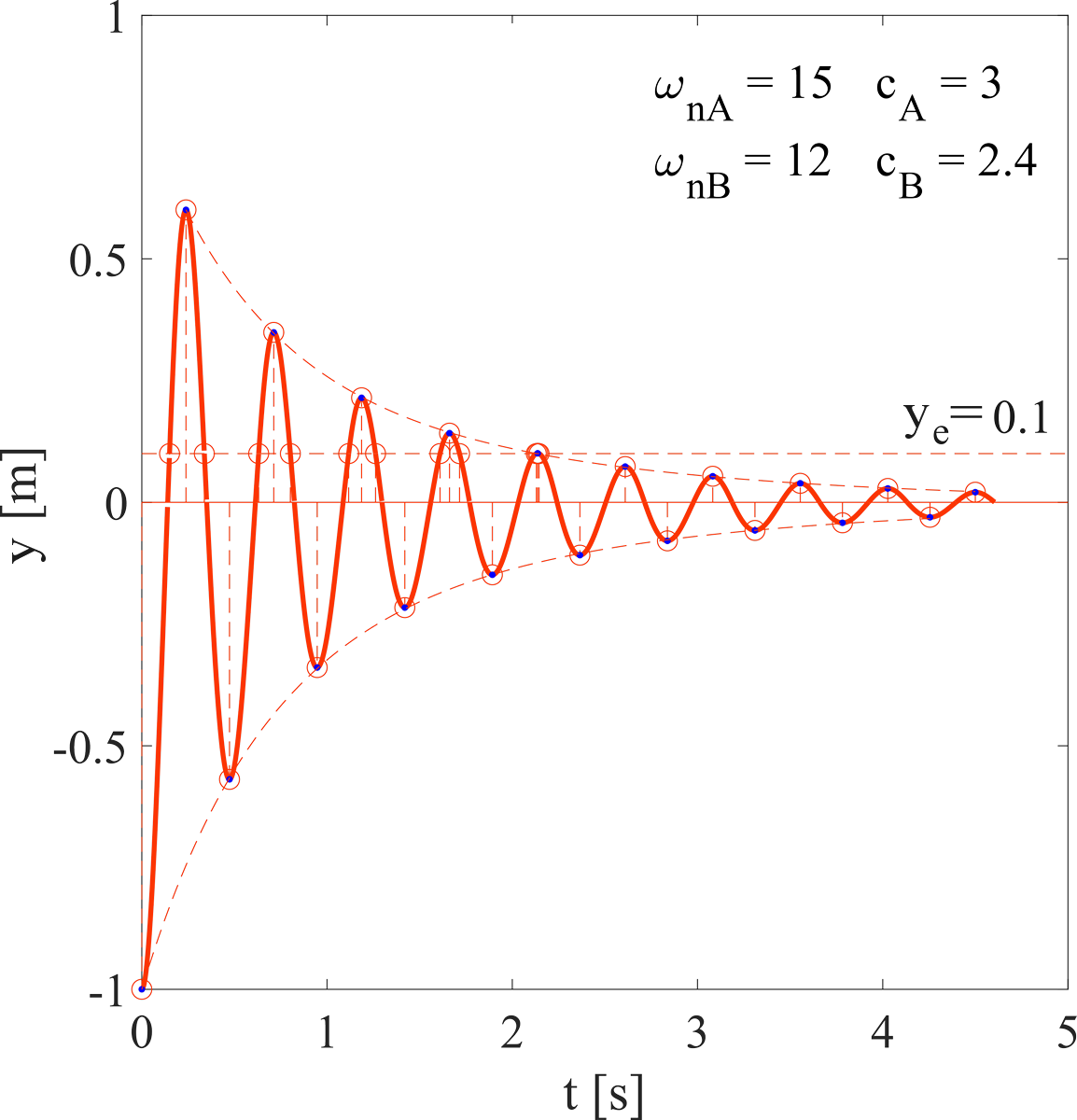}
 & \includegraphics[width=2.7in]{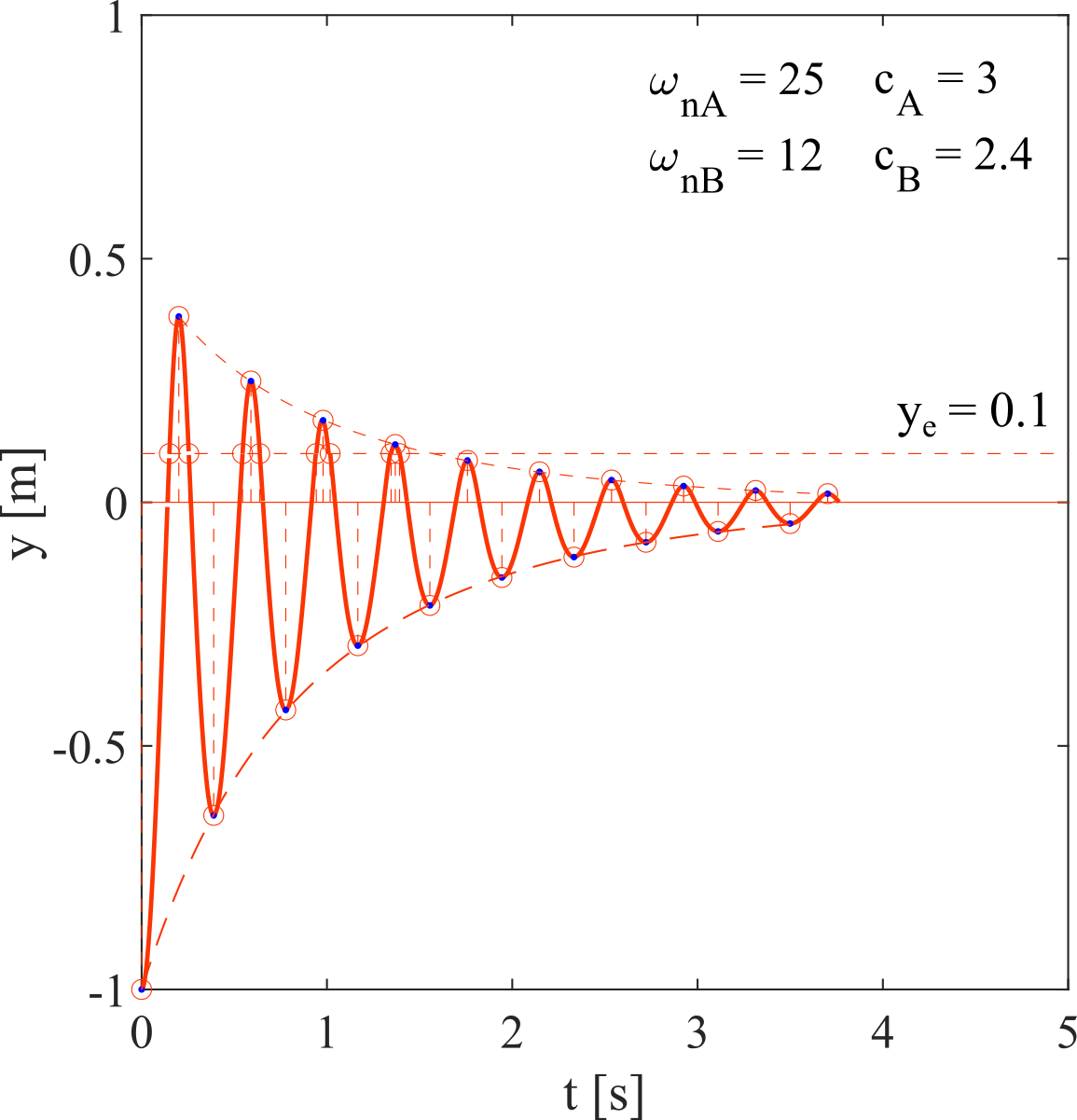}\\
 (c) & (d)\\
\end{tabular}
\caption{Logarithmic decrement asymmetry exhibited when partial engagement at $y_\text{e}>0$ is present in a simplified system of two coupled damped oscillators. (a) A schematic of asymmetric spring mass damper system simulating the biomimetic metastructure. The offset mimics initial scale angle after which engagement occurs. The engagement adds additional stiffness $\Delta\omega$ and additional damping $\Delta c$ (b) Damped oscillations with no offset, (c) with offset, (d) with offset and higher stiffness on scales side than (c).}
\label{Asymmetry}
\end{figure*}

%This is in line with the overall effect of higher $\eta$ in increasing the anisotropy of vibration itself. 

In order to gain better physical insight into the system we probe the fundamentals using an asymmetric spring mass damper system (SMD) that is damped more on one side, Fig.~\ref{Asymmetry}. This system can be integrated to obtain analytical closed-form expressions for logarithmic damping ratio (See Supplementary Material).

The symmetry breaking of this system can result from two different sources - scales sliding on one side, and delayed engagement of scales on the other side only due to initial scale angle. Damping phase maps show that although introduction of asymmetry (in one direction) is sufficient to cause overall asymmetry of the system, the logarithmic decrement is still the same for both sides of damped oscillation. Thus damping on just one side is not sufficient to cause a major change in symmetry in logarithmic damping ratio. The symmetry of the SMD system can be further broken if we introduce an engagement asymmetry to mimic initial angle $\theta_0$. The asymmetry in stiffness in our SMD model is addressed by letting that damping start at $y=0$ in the domain $y<0$, and at an arbitrary $y_\text{e}>0$ in the domain $y>0$, Fig.~\ref{Asymmetry} (a). We vary the natural frequencies $\omega|_{y>y_\text{e}}$ and $\omega|_{y<0}$ while keeping the damping coefficients $c|_{y>0}$ and $c|_{y<0}$ constant, so that damping ratios $\zeta|_{y>0}$ and $\zeta|_{y<0}$ change together with the respective natural frequencies. These parameters are defined to approximate our architectured system – high natural frequencies indicate higher stiffness, where higher damping coefficient is meant to simulate higher stiffness, whereas $y_\text{e}$–-the offset--simulates the initial angle. In Fig. 4 (b) we plot the effect of scales with no offset. We clearly see that the lack of offset results in negligible difference in damping between the scales and the plain side. However, as soon as offset is added we see that a visible asymmetry in damping emerges. The asymmetry in damping coefficient increases as the contrast between the two sides increases. In Fig. 4 (c), we plot the effect of different damping coefficients with a given offset, and in Fig 4 (d), we increase the ratio of the stiffnesses while keeping the damping coefficients the same. This would be a comparison between systems with a different overlap ratio.  We find that higher overlap ratio clearly accentuates the asymmetry in logarithmic damping, ceteris paribus. This confirms the trends from phase map Fig. 3 (c), (d). Here, we compare two asymmetric SMD systems, one with no offset and another with offset and find that addition of offset, accentuates asymmetry, confirming our findings in Fig 3 (a)-(b).
In addition to free vibration, we also consider dissipative effects in forced oscillations. We quantify dissipation using "Specific Damping Capacity (\textit{SDC})", which measures a material's ability to dissipate elastic strain energy through a mechanical vibration motion \cite{zhang1993documentation,rao2019vibration}. The \textit{SDC} can be defined as follows:

\begin{equation}
  SDC = \frac{\mathrm{Dissipated \hspace{3pt} Energy \hspace{3pt} per \hspace{3pt} Steady \hspace{3pt} State \hspace{3pt} Cycle} \hspace{3pt} (\Delta U)}{\mathrm{Maximum \hspace{3pt} Stored \hspace{3pt} Energy \hspace{3pt} (U)}}.
  \label{Eq8}    
\end{equation}
\vspace{1pt}

% According to the work-energy balance in a steady state forced vibration, the dissipated energy per cycle is equal to the work done by the applied force. Also, for this scaled-covered system, the maximum stored energy is the maximum of the summation of the beam's strain energy and the strain energy due to the scales rotation. 
These dissipated and stored energies can be calculated numerically through a computational model.

\begin{figure*}
 \centering
 \begin{tabular}{cc}
 \includegraphics[width=3in]{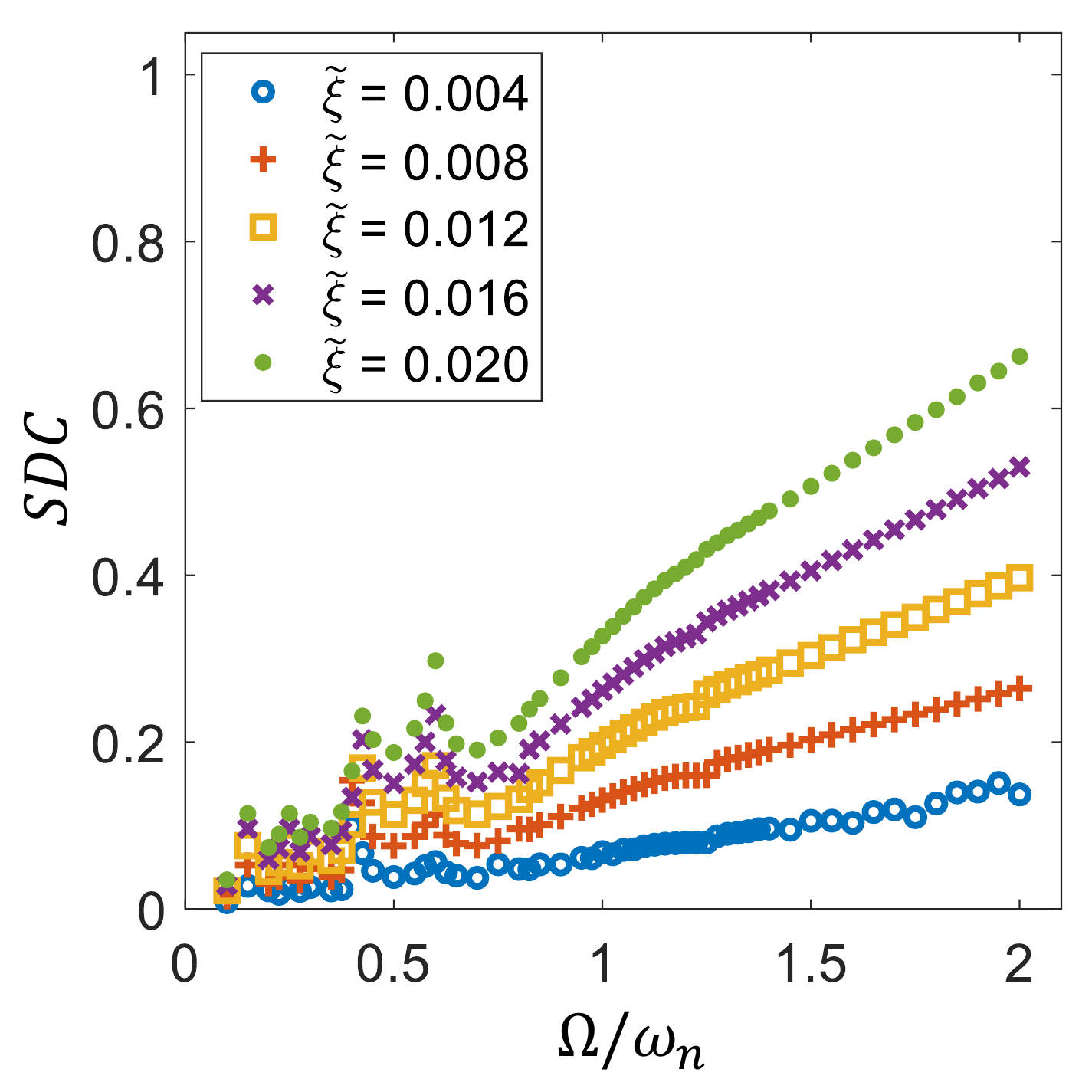} 
 & \includegraphics[width=3in]{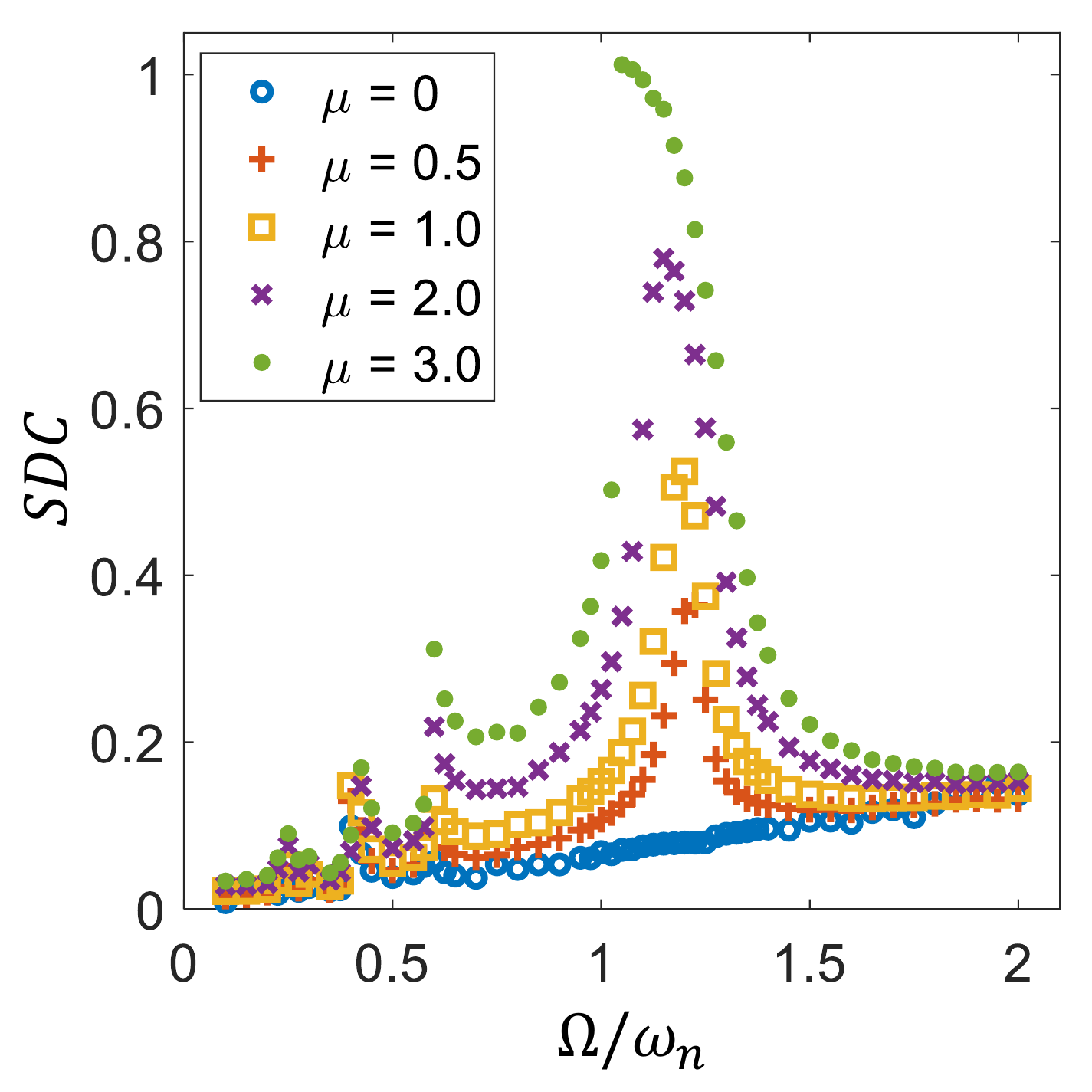} \\
 (a) & (b) \\
 % \includegraphics[width=3in]{Fig5(c).pdf} 
 % & \includegraphics[width=3in]{Fig5(d).pdf} \\
 % (c) & (d) \\
\end{tabular}
 \caption{%(a) Specific Damping Capacity (\textit{SDC}) for various $\tilde{\xi}$, with $\mu=0$, $\eta=5$, and $\theta_0=5^\circ$. (b) Specific Damping Capacity (\textit{SDC}) for various $\mu$, with $\tilde{\xi}=0.004$, $\eta=5$, and $\theta_0=5^\circ$.
 (a) Variation of Specific damping coefficient (SDC) with frequency for forced vibration for various material viscosity parameters, friction is absent. (b) Variation of Specific damping coefficient (SDC) with frequency for forced vibration for various coefficients of friction. Material viscosity is negligible (<0.005)}
 \label{Fig5}
\end{figure*}

% To further distinguish the material vs the geometrical sources of dissipation, we take recourse to the following Fig.~\ref{Fig5}. 
Here, in Fig.~\ref{Fig5} (a), we plot the specific damping capacity from material sources $\bar \xi$ and find that it increases (barring a few peaks at sub-harmonic frequencies due to complex nature of oscillations) as the frequency increases. This is a traditional viscous damping response. %as also reflected in the force-amplitude plot, Fig.~\ref{Fig5} (b) for various values of $\bar \xi$. 
In contrast, the effect of inter-scale friction is dramatically different. The specific damping capacity, Fig.~\ref{Fig5} (b) shows a pronounced and sharp increase near resonance with higher peaks corresponding to higher friction. After the resonance, the damping begins to decrease sharply. The overall reason for this behavior is due to lack of rate-dependence of the frictional component of the force. As the amplitude of the vibration decreases post resonance, so does the work done by friction. %The amplitude-frequency curve for friction is shown in  Fig.~\ref{Fig5} (d), which shows a much more pronounced nonlinearity than the one with only material non-linearity. 

 % \caption{(a) . (b) . (c) . (d) .}

In conclusion, we find that although the viscoelastic and frictional sources of dissipation are two apparently similar sources of damping in a biomimetic scale architectured substrate, on closer scrutiny they are quite different. Their effects on displacement, damping asymmetry, and specific damping markedly diverge. The geometry-material interplay is investigated for the first time. Real world polymers exhibit far more complexity in their material behavior. In linear regime they are often represented by a combination of Kelvin-Voigt elements. We aim to study such complexities in later iterations of this study. In spite of this limitation, the current findings have wide implications in the design of fish scale like smart skins and appendages for soft robotics, tailored prosthetic applications. 

\begin{acknowledgments}
This work was supported by the United States National Science Foundation’s
Civil, Mechanical, and Manufacturing Innovation, Grant No. 2028338.
\end{acknowledgments}

% \appendix

% \section{Appendixes}

% To start the appendixes, use the \verb+\appendix+ command.
% This signals that all following section commands refer to appendixes
% instead of regular sections. Therefore, the \verb+\appendix+ command
% should be used only once---to set up the section commands to act as
% appendixes. Thereafter normal section commands are used. The heading
% for a section can be left empty. For example,
% \begin{verbatim}
% \appendix
% \section{}
% \end{verbatim}
% will produce an appendix heading that says ``APPENDIX A'' and
% \begin{verbatim}
% \appendix
% \section{Background}
% \end{verbatim}
% will produce an appendix heading that says ``APPENDIX A: BACKGROUND''
% (note that the colon is set automatically).

% If there is only one appendix, then the letter ``A'' should not
% appear. This is suppressed by using the star version of the appendix
% command (\verb+\appendix*+ in the place of \verb+\appendix+).

\appendix

%\nocite{*}
\bibliography{aipsamp}% Produces the bibliography via BibTeX.

\end{document}